\documentclass[a4]{article}

\usepackage{amsfonts}
\usepackage{latexsym}
\usepackage{xypic}

\newcommand{\notext}[1]{\mbox{}}
\newcommand{\sem}[1]{[\! [ #1 ]\! ]}
\newcommand{\mysem}[1]{[\! [ #1 ]\! ]}
\newcommand{\equiva}[0]{=\!\!\! |\models}
\newcommand{\tupel}[1]{\langle #1 \rangle}
\newcommand{\verz}[1]{\{ #1\} }
\newcommand{\pow}[1]{\mbox{\sc pow}( #1 )}

\newcommand{\infer}[0]{\mbox{\it infer}}
\newcommand{\error}[0]{\mbox{\sc error}}
\newcommand{\states}[0]{\mbox{\sc states}}

\newcommand{\weg}[0]{\mbox{\sc drop}}

\newcommand{\inc}[0]{\emptyset}
\newcommand{\cons}[0]{\mbox{\sc cons}}

\newcommand{\csp}[0]{{\cal C}}

\newtheorem{definition}{Definition}[section]
\newtheorem{proposition}{Proposition}[section]
\newtheorem{theorem}{Theorem}
\newtheorem{lemma}{Lemma}
\newtheorem{exa}{Example}

\newenvironment{proof}{\mbox{}\linebreak {\bf Proof}:$\;$}{\hfill $\Box$\\}

\begin{document}
\title{Sound search in a denotational semantics for first order logic}
\author{C.F.M.~Vermeulen}
\maketitle
\begin{abstract}
In this paper we adapt the definitions and results from Apt and Vermeulen
in \cite{lpar} to include
important ideas about {\it search} and {\it choice} into the system. We 
give motivating 
examples. Then we set up denotational semantics for first order logic 
along the lines of Apt \cite{Apt00} and  Apt and Vermeulen \cite{lpar}. 
The semantic universe 
includes states that consist of two components: a substitution, which can be 
seen as the computed answer; and a constraint satisfaction problem, which can 
be seen as the residue of the original problem, yet to be handled by constraint 
programming.
In the set up the interaction between these components is regulated by
an operator called {\it infer}.
In this paper we regard
{\it infer} as an operator on sets of states to enable us
to analyze ideas about {\it search} among states and {\it choice}
between states.

The precise adaptations of definitions and results are able to deal with 
the examples and we show 
that given several reasonable conditions, the new definitions ensure soundness
of the system with respect to the standard interpretation of first order logic.
In this way the `reasonable conditions' can be read as {\it conditions 
for sound search}. We indicate briefly how to investigate {\it efficiency} of
search in future research.
\end{abstract}

\section{Introduction}\label{intro}

The motivating examples for this paper are those examples in 
constraint programming where a constraint satisfaction problem (CSP) can be
handled by distinguishing cases. Let's look at two such examples: first 
a rather trivial one, mainly for illustrative purposes; then a more realistic 
one that is taken from the current literature on constraint programming. 
All notation will be properly explained later on, but we already employ 
some of it
in the presentation of the examples. 

First consider the constraint satisfaction problem $x^2=1$ in a situation 
in which none of the values of the variables have been computed yet. We 
write this:
\begin{quote}
$\csp_0=\tupel{x^2=1\; ;\; \epsilon}$
\end{quote}
where $\epsilon$ is the empty substitution.
We may be able to feed this problem to a constraint propagation tool that
transfers it into the equivalent form:
\begin{quote}
$\csp_1=\tupel{x=1\;\vee\; x=-1\; ;\; \epsilon}$.
\end{quote}
In \cite{lpar} Apt and Vermeulen show how to formalize such a step.
The preservation 
of equivalence in the transition from $\csp_0$ 
to $\csp_1$ is vital. It is essential for the {\it soundness} result
in \cite{lpar}. Such a soundness result is the way to check that the 
computation
steps in the system are `all right' according to first order logic.

Now we have a CSP $\csp_1$ that is {\it disjunctive}. So, it makes sense to
distinguishes two cases:
\begin{quote}
$\csp_2=\tupel{x=1\; ;\; \epsilon}$ and
$\csp_3=\tupel{x=-1\; ;\; \epsilon}$,
\end{quote}
and proceed by distinguishing these two cases and compute
the value of the variable $x$ in each case:
\begin{quote}
$\csp_4=\tupel{\top\; ;\; \verz{x/{\bf 1}}}$ and
$\csp_5=\tupel{\top\; ;\; \verz{x/-{\bf 1}}}$.
\end{quote}
But by splitting things up we lose equivalence: neither $\csp_2$ nor $\csp_3$
are equivalent to the original $\csp_0$. And this loss of equivalence
frustrates the soundness result from \cite{lpar}.

Another interesting example 
is the search for a suitable value for a variable $x$ 
in a domain ${\cal D}_x$. Such a search could be organized by following some 
way of {\it ranking} the values $a\in{\cal D}_x$. Discussion of variable
ranking can be found, for example, in Milano and Van Hoeve \cite{WJ02}.
They employ the ranking for distinguishing among the values ${\cal D}_x$
for $x$ a subset
of promising values, called ${\cal D}_{good}\subset{\cal D}_x$. Then the 
search for a solution to a CSP can be speeded up by splitting up the original
CSP, first considering the {\it good} values in ${\cal D}_{good}$
and considering the less promising 
part of $D_x$ later. Such tricks can result in a notable speed up for several
tasks in constraint programming. For example, it is clear that an
(in-)consistency check  for a CSP, can benefit from such a distinction of likely 
and un-likely values. But the distinction does not satisfy the equivalence 
condition from \cite{lpar}. Hence there is no general result to guarantee the
soundness of the search strategy.

So, the situation is that we would like to be able to analyze search strategies 
and case distinction with the level of generality that is achieved by
Apt and Vermeulen in \cite{lpar}.
But in the semantics of \cite{lpar} methods for 
search and distinction of cases
cannot readily be modeled, as it relies on the preservation of
{\it equivalence} in the transition from one constraint satisfaction 
problem to the next. Therefore we propose a different notion of 
preservation of equivalence here, called {\it pointwise equivalence}, and
show how with this new notion of equivalence 
the system in \cite{lpar} can be adapted
to suit the analysis of search strategies and case distinctions.
In particular, we will show how this adaptation can be made without losing 
the crucial soundness properties of the semantics, so that the intended 
connection with the standard interpretation of first order logic remains
intact.

\section{Constraint propagation in first order logic}
\label{recol}

We repeat the facts from \cite{lpar} and the notation introduced there.

\label{sub:prelim}

Let's assume that an algebra ${\cal J}$ is given over which we want to perform
computations. This can be for example: the standard algebra for the language 
of arithmetic, in case we want to find solutions to equations or systems 
of equations; the algebra of terms over a first order language, in case we 
want to compute unifiers of terms; etc.

In each case the basic ingredient of the semantic universe will be the set of 
states, $\states$. States come
in two kinds. First we have an $\error$ state, which remains unanalyzed.
All other states consist of two components: one component is a constraint 
satisfaction problem  $\csp$, the other a substitution $\theta$. 
Such a state is then written 
$\tupel{\csp ;\theta}$. As always, a substitution 
$\theta$ is a mapping from variables to terms. It assigns a term 
$x\theta$ to each variable $x$, but there are only finitely many variables 
for which $x\neq x\theta$. These variables form $dom(\theta )$, the 
domain of $\theta$. 
The application of a substitution $\theta$ to a term $t$, written 
$t\theta$, is defined as usual. We denote the empty substitution by
$\epsilon$. There is another convenient
notation concerning substitutions: we write
$\hat\theta$ for the conjunction 
$\bigwedge\verz{x=x\theta:\; x\in dom(\theta )}$.

However our notion of substitution deviates from the 
usual notion of substitution in that  the terms we assign to a variable are 
always partially evaluated in the intended algebra ${\cal I}$. This trick was 
developed and motivated  by Apt in \cite{Apt00}. For example, if 
${\cal I}$ is the standard algebra for the language of arithmetic, and we 
find that $x=5$, then the substitution will set $\verz{x/ {\bf 5}}$, i.e. we 
assign the integer ${\bf 5}$ to $x$ rather than the term $5$. This strategy of 
evaluating as much as possible is then extended systematically. So, 
\begin{quote}
if 
$x\theta = {\bf 4}$ and $y\theta = z$, then $(x+y)\theta = {\bf 4} + z$.
\end{quote} We can 
only compute the value for $x+y$ partially. But,
\begin{quote}
if
$x\theta ={\bf 4}$ and $y\theta ={\bf 5}$, then $(x+y)\theta = {\bf 9}$.
\end{quote}
Now we can already compute the value of $x+y\in {\cal I}$ completely.
We refer to Apt's \cite{Apt00} for more details on this 
trick for partial evaluation.
Its main advantage is that we can now already use some special properties of the 
algebra ${\cal I}$ during the computation.

For the second example mentioned above, where ${\cal I}$ is an 
algebra of terms, the trick for partial evaluation does not
make a difference: the
partially evaluated substitutions and the usual notion of substitution give 
exactly the same results. So, we find the standard notion of substitution
of logic programming as a special case.

A constraint satisfaction problem (CSP) $\csp$, 
simply is a finite set of formulas of first order logic. In many applications
there are extra requirements on the syntactic form of a CSP,
but for now we keep things as general as possible. $\bot$ is a special
formula which is always false. We also write $\csp$ for $\bigwedge \csp$,
the conjunction over the formulas in $\csp$.
For a set of states $S$ we write $\bigvee S$ for the disjunction of all the
formulas $\csp\wedge\hat\theta$ (for $\tupel{\csp ;\theta}\in S$).

Throughout the paper we try to limit the number of brackets 
and braces
as much as possible. In particular, for a finite set $\verz{A_1,\ldots,A_n}$
we often write $A_1, \ldots,A_n$. Also, 
we write $\infer\tupel{\csp ;\theta }$ instead  of 
$\infer(\tupel{\csp  ;\theta})$, etc.

For the treatment of local variables we introduce a mapping $\mbox{\sc drop}_u$ 
(for each variable $u$). First we define {\it DROP}$_u$, a mapping 
on substitutions:\\

\begin{definition}
\mbox{}\\

\begin{tabular}{ll}
$u {\it DROP}_u(\theta )$ $=\; u$\\
$x {\it DROP}_u(\theta )$ $=\; x\theta$ \mbox{for all other variables $x$}\\
\end{tabular}\\

\end{definition}

So, ${\it dom} ({\it DROP}_u(\theta ))={\it dom} (\theta ) -\verz{u}$.

We write $\csp (u)$ for the part of $\csp$ in which $u$ really occurs. 
Then we can define $\mbox{\sc drop}_u$, a mapping on $\states$.

\begin{definition}
\mbox{}\\

\begin{tabular}{llll}
$\;\mbox{\sc drop}_u\tupel{\csp  ;\eta}$ & $=$&$ \tupel{\csp  ;
        \mbox{\it DROP}_u (\eta ) }$ & if $\csp (u)=\emptyset\;\;$\\[2mm]
$\;\mbox{\sc drop}_u\tupel{\csp  ;\eta}$ & $=$&
        $ \langle\;\exists u\; (u=u\eta \;\wedge\; {\bf y}={\bf y}\eta
        \;\wedge\; \csp (u)),$&\\
$\;$ & &        \hspace{2cm}$\; \csp -\csp (u);\;\;\;\;\;\;
                \mbox{\it DROP}_u (\eta )\rangle$
        & if $\csp (u)\neq\emptyset\;\;$\\[2mm]
$\;\mbox{\sc drop}_u \error $ & $=$ & $\error$ &\\
$\;$&&&\\
\end{tabular}

Here ${\bf y}$ denotes the sequence of variables $y_1,\ldots ,y_n$ such that
$u\in y_i\eta$.
\end{definition}

This shows that \mbox{\sc drop}$_u$ removes $u$ from the domain of
the assignment $\theta$ and existentially quantifies the occurrences of $u$
in the CSP $\csp$.\\

In the definition of the denotational semantics we meet a parameter called
$\infer$. This is the crucial parameter in our story. $\infer$ maps
sets of $\states$ to sets of $\states$. It can be instantiated to cover all kinds 
of constraint propagation (cf. Apt and Vermeulen \cite{lpar}). Important 
examples to keep in mind are:
the case where ${\cal I}$ is a term algebra and the constraint propagation tool 
performs unification; the case where ${\cal I}$ is the standard algebra of 
arithmetic and the constraint propagation tool can compute answers for certain
types of equations very efficiently. In the first case we could find, for example,
that:
\begin{quote}
$\tupel{\top ;\verz{x/z ,y/z}}\in {\it infer}(\tupel{f(x) = f(y)\; ,\; \epsilon})$.
\end{quote}
I.e., the {\it infer} operation computes the unifying substitution 
$\verz{x/z ,y/z}$.
In the second example we may have a constraint propagation mechanism
that solves certain quadratic equations over the integers and find that
\begin{quote}
$\tupel{x=1\vee x=-1;\; \epsilon }\in {\it infer}(\tupel{x^2=1\; ,\; \epsilon})$.
\end{quote}

This type of constraint propagation could already be covered by Apt and Vermeulen
in \cite{lpar}.
Here we also incorporate the analysis of search strategies
within sets of $\states$ and the generation of subproblems from $\states$.
Then it can actually happen that we find:
\begin{quote}
$\tupel{\top ;\; \verz{x/{\bf 1}} }\in {\it infer}(\tupel{x^2=1\; ,\; \epsilon})$
and

$\tupel{\top ;\; \verz{x/-{\bf 1}} }\in {\it infer}(\tupel{x^2=1\; ,\; \epsilon})$.
\end{quote}
So, {\it infer} will be able to come up with two essentially
{\it distinct} computed answers. This was not allowed in \cite{lpar}.

Below we will discuss natural conditions on the 
$\infer$ operator that guarantee that the computations performed 
by ${\it infer}$
respect first order logic. But for the definition we do not have to worry 
about these conditions yet.\\

We can now present our denotational semantics for first order logic 
in which the $\infer$ mapping is a parameter, as explained above. 
By having the $\infer$ parameter 
we obtain general results, that apply uniformly to various 
forms of constraint store management, search and case distinction.

We define the mapping 
$\mysem{\phi}:\; \pow{\states }\rightarrow\pow{\states}$, using 
postfix notation. The definition is presented {\it pointwise}, for singleton sets 
$\verz{\tupel{\csp;\theta}}$. Then the general case is fixed by the equation
$S\mysem{\phi}=
\bigcup\verz{\tupel{\csp;\theta}\mysem{\phi}:\;
\tupel{\csp;\theta}\in S}$.\\

A final bit of notation is $\mbox{\cons}^+(S)$, which is the subset of $S$ 
that contains exactly those $\states$ that are not inconsistent. So:
$\mbox{\sc cons}^+(S)=\verz{\error:\; \error\in S}\cup\verz{\tupel{\csp;\theta}
\in S:\; \not \models_{\cal I} \neg (\csp\wedge\hat\theta)}$.
We will use $\mbox{\sc cons}(S)$ for $\mbox{\sc cons}^+(S)-\verz{\error}$.

\begin{definition}
\mbox{}\\

\begin{tabular}{|ll|}
\hline
&\\ $\;\tupel{\csp  ;\theta}\mysem{A}\; $&
	$=\; \infer \tupel{\csp ,{A}  ;\theta}$
        \hspace{.8cm} for an atomic formula $A$
\\[2mm] $\;\tupel{\csp  ;\theta}\mysem{\phi_1\vee\phi_2}\;$&$ =\; 
        \tupel{\csp  ;\theta}\mysem{\phi_1}\; \cup\; 
        \tupel{\csp  ;\theta}\mysem{\phi_2}$
\\[2mm] $\;\tupel{\csp  ;\theta}\mysem{\phi_1\wedge\phi_2}\;$&$ =\;
        (\tupel{\csp ;\theta}\mysem{\phi_1})\mysem{\phi_2}$
\\[2mm] $\;\tupel{\csp  ;\theta}\mysem{\neg\phi}\;$&$ =
        \left\{
        \begin{tabular}{ll}
        $\infer\tupel{\csp  ;\theta}$ &
        \mbox{if } $\cons^+ (\tupel{\csp  ;\theta}\mysem{\phi})=\emptyset$\\
        
        $\emptyset$ &
        \mbox{if }$\tupel{\csp ' ;\theta '}\in
                \cons(\tupel{\csp  ;\theta}\mysem{\phi})$
        \mbox{ for}\\ &
        \mbox{some $\tupel{\csp ' ;\theta '}$ equivalent to
        $\tupel{\csp  ;\theta}$} \\[2mm]
        $\infer\tupel{\csp ;{\neg\phi}  ;\theta}$ &
        \mbox{otherwise}
        \end{tabular}
        \right.$\\[2mm]
$\;\tupel{\csp  ;\theta}\mysem{\exists x\; \phi}\;$&$ =\;
%
        \bigcup_{\sigma}\verz{\infer{\;\mbox{\sc drop}_u
                (\sigma )}}$

        where, for some fresh $u$,\\
        & \hspace{1.6cm} $\sigma$ ranges over $\cons^+ (
        \tupel{\csp  ;\theta}\mysem{\phi\verz{x/u}})$ 
\\[2mm] $\;\error \mysem{\phi} \;$&$=\; \verz{\error}$  
		\mbox{ for all $\phi$}\\
&\\ \hline
\end{tabular}

\end{definition}

The definition relies heavily on the notation that was introduced before. But
it is still quite easy to see what goes on. The atomic formulas are
handled by means of the 
$\infer$ mapping. Then, disjunction is interpreted as nondeterministic
choice,
and conjunction as sequential composition. For existential quantification
we use the $\weg _u$ mapping (for a fresh variable $u$). The $\error$ clause
says that there is no recovery from $\error$. In the case for negation,
three contingencies are present: first, the case where $\phi$ is 
inconsistent (${\cons}^+(\tupel{\csp ;\theta}\sem{\phi}=\emptyset$). Then we
continue with the input state $\tupel{\csp  ;\theta}$.
Secondly, the case where $\phi$ is already true in (a state 
equivalent to)
the input state. Then we conclude that $\neg\phi$ yields 
inconsistence, i.e., we get $\emptyset$. Finally, we add $\neg\phi$
to the constraint store $\csp$ if it is impossible at this point to reach
a decision about  the status of $\neg\phi$.

Here we have made the choice to use $\emptyset$ for {\it inconsistency} or 
{\it falsehood}. Strictly speaking there is something arbitrary about this 
choice. Any set $S$ such that ${\cons}^+ (S)=\emptyset$ would have done equally 
well. We will see this also later on, when we discuss the formulation of the 
soundness theorem and the inconsistency condition (3) on {\it infer}.

\label{sec:sound}

Next we show that the denotational
semantics with the $\infer$ parameter is sound. This 
amounts to two things:
{\it 1.}  successful
computations of $\phi$ result in states in which $\phi$ holds;
{\it 2.}  if no 
successful computation of $\phi$ exists, $\phi$ is false in 
the initial state.
So, the soundness result show that the denotational semantics 
respects the standard semantics for first order logic.

\section{Conditions on propagation and search}
\label{cond}

In \cite{lpar} we formulated natural conditions on the instantiations of the 
{\it infer} operation. The effect of these conditions was that we could 
prove the {\it soundness} of the semantics, i.e, we could show that for each 
setting
of the {\it infer} mapping that satisfies the conditions, we get a denotational
semantics for first order logic that
respects the standard interpretation of first order logic.
As in \cite{lpar} we use {\it infer} to analyze constraint propagation,
the conditions can be seen as conditions for sound propagation.

The conditions from \cite {lpar} are as follows:\footnote{ $\phi \equiva \psi$ is 
shorthand for: both $\phi\models\psi$ and $\psi\models\phi$.} 

\begin{description}
\item[(1)] If $\tupel{\csp ' ,\theta}\in \infer(\tupel{\csp ,\theta} )$, then
	$\csp\wedge\hat\theta\;\equiva\; \csp '\wedge\hat{\theta '}$
	\hfill{(equivalence)}
\item[(2)] If $\tupel{\csp ' ,\theta '}\in \infer(\tupel{\csp ,\theta} )$, then
	also $\tupel{\csp '_v,\theta '_v} \in \infer (\tupel{\csp ,\theta})$,
	where $\tupel{\csp '_v,\theta '_v}$ is obtained from 
	$\tupel{\csp ',\theta '}$ by systematically replacing all occurrences
	of $u$ by $v$ for a variable $u$ that is fresh w.r.t. 
	$\tupel{\csp ,\theta}$ and a variable $v$ that is fresh w.r.t. both
	$\tupel{\csp ,\theta}$ and $\tupel{\csp ',\theta '}$ \hfill(alphabetic
	variation)
\item[(3)] If $\infer(\tupel{\csp ,\theta})=\emptyset$, then 
	$\csp\theta\models_{\cal I}\bot$. \hfill{(inconsistency)}
\item[(4)] $\infer(\error )=\verz{\error}$. \hfill{(error)}
\end{description}

Condition (1) is an equivalence condition: it insists on preservation
of logical equivalence by the operation {\it infer}. Condition (2)
is awkward to read, but it turns out to be the appropriate way of saying that 
{\it infer} should not depend on specific choices of fresh variables.
Condition (3) insists on the preservation of inconsistency by {\it infer}
and condition (4) is about the propagation of $\error$.\footnote{Please 
recall the remark about using $\emptyset$ as the 
one-and-only inconsistency indicator. If we do not follow this policy, we can
weaken the condition in (3) to: ${\cons}^+(\infer(\tupel{\csp ,\theta}))
=\emptyset$.}

In this paper we have lifted the {\it infer} operation to an operation
on {\it sets} of ${\sc states}$. So, we also have to lift these
conditions in an appropriate way.
Fortunately, for conditions (2)-(4) we do not have to change 
anything.
But we do need 
to adapt the equivalence condition (1), as we have seen that it is not satisfied
by the examples of case distinctions from section \ref{intro} that we want to 
analyze.
The obvious way to adapt the equivalence condition on $\infer$ is perhaps:

\begin{definition}
Set Equivalence: $\bigvee\infer(S)\;\equiva\;\bigvee S$
\end{definition}
where $\phi\;\equiva\;\psi$ stands for: $\phi\models\psi$ and 
$\psi\models\phi$
and $\bigvee S$ is notation for: $\bigvee\verz{{\cal C}\wedge\hat{\theta}:\;
\tupel{{\cal C},\theta}\in S}$.

This condition allows us to split states in an infer step, as required in
the examples. It allows for 
\begin{quote}
${\it infer}(\tupel{x=1\vee x=-1 ;\epsilon}=\verz{\tupel{x=1 ;\epsilon},
\tupel{x=-1 ;\epsilon}}$,
\end{quote}
as required. But 
it also allows us to re-group states in a confusing way. For example,
if $\tupel{x=1;\epsilon},\; \tupel{x=-1;\epsilon}\in S$, the set equivalence
condition allows us to re-group this and have 
$\tupel{x=1\;\vee\; x=-1;\epsilon}\in\infer(S)$.
If we consider an example with three options, 
$\tupel{x=1\vee x=2\vee x=3;\epsilon}$ for instance, we can see even more 
confusing forms of grouping and re-grouping.
More generally, this condition gives us no control over the origin of states in 
$\infer (S)$. It does not tell us where a particular state 
$\tupel{{\cal C};\theta}\in\infer (S)$ is coming from.

This is not the sort of search we are trying to cover and we will see that
it messes things up in the soundness proof as well.

Instead, we opt for pointwise equivalence:

\begin{definition}
Pointwise Equivalence: $\bigvee\infer(\tupel{{\cal C};\theta})\;\equiva\;
{\cal C}\wedge\hat\theta$ for each $S$ and each $\tupel{{\cal C};\theta}\in S$
\end{definition}

Now we can relate individual members of $\infer (S)$ to {\it ancestors}
in $S$. And the {\it pointwise equivalence} condition makes sure that
each state in $S$ is equivalent to the set of its descendants in 
${\it infer}S$.
The definitions allow for one state to have several ancestors, but each of 
these ancestors has to be able to account for its descendants by itself.
This way re-grouping as in the example above is no longer allowed. This is
entirely compatible with our motivation and it will
make the soundness proof run smoothly.

The following property relates the two conditions on $\infer$:

\begin{definition}
Continuity: $\infer (S)\; =\; 
\bigcup\verz{\infer(\tupel{{\cal C};\theta}):\; \tupel{{\cal C};\theta}\in S}$
\end{definition}

\begin{proposition}Assume that $\infer$ satisfies the {\it continuity} 
condition.
Then set equivalence and pointwise equivalence are equivalent.
\end{proposition}

\begin{proof}
Note that the proposition is about the equivalence of two equivalence 
conditions. First we check that set equivalence implies pointwise
equivalence.

\begin{itemize}
\item We apply set equivalence to the one element set
$\verz{\tupel{\csp ;\theta}}$ for
$\tupel{\csp ;\theta}\in S$. This gives: $\bigvee\infer \tupel{\csp ;\theta}\;
\equiva\; \csp\wedge\hat\theta$, as required. (Note that we do not need
{\it continuity}.)
\end{itemize}

Next we assume pointwise equivalence and check set equivalence.
This is an exercise in handling disjunctions in propositional logic.

\begin{itemize}
\item Consider $\bigvee S$. This is a disjunction of formulas 
of form $\csp\wedge\hat
\theta$ (for $\tupel{\csp ;\theta}\in S$.)
So, to establish $\bigvee S\models\bigvee\infer S$, it suffices to check that 
$\csp\wedge\hat\theta \models \bigvee\infer S$, for each 
$\tupel{\csp ;\theta}\in S$.
From pointwise equivalence we readily obtain:
$\csp\wedge\hat\theta \models\bigvee\infer \tupel{\csp ;\theta}$.
Now, by continuity we get: $\bigvee\infer \tupel{\csp ;\theta}\models
\bigvee\infer S$. Jointly this gives: $\csp\wedge\hat\theta \models
\bigvee\infer S$, as required.
\item Next consider $\bigvee\infer S$. We need to establish:
$\bigvee\infer S\models\bigvee S$. By continuity we know that
$\bigvee\infer S$ is the disjunction of all the 
$\bigvee\infer \tupel{\csp ;\theta}$ (for $\tupel{\csp ;\theta}\in S$). 
So, it suffices to check that each 
of these smaller disjunctions entails $\bigvee S$. Pointwise equivalence
ensures that: $\bigvee\infer \tupel{\csp ;\theta}\models 
\csp\wedge\hat\theta$ and hence: $\bigvee\infer \tupel{\csp ;\theta}\models
\bigvee S$ follows simply by propositional logic.
\end{itemize}
\end{proof}

Below we will always assume pointwise equivalence for $\infer$.
Note that the formulation in this paper makes condition (3) a consequence of
equivalence. Below we will only refer to condition (3) separately if this 
adds anything to the readability of the proofs.

\section{Sound propagation and search}
\label{sound}

We start by stating the soundness claim and the preservation lemma in the new 
setting:\footnote{In this section we insist on mentioning ${\cal I}$ all the
time to remind us that we are looking at the choice of values from ${\cal I}$.}

\begin{theorem}[Soundness]
Let $S$, $\tupel{\csp,\theta}\in S$, $\phi$ be given. Then we have:

\begin{enumerate}
\item $\bigvee S\sem{\phi}\models_{\cal I}\phi$
%
\item If ${\cons}^+(S\sem{\phi}) = \inc$ ,
        then $\bigvee S\models_{\cal I}\neg\phi$.
\end{enumerate}
\end{theorem}

Here $\inc$ is a sign of {\it inconsistency} or {\it falsehood}: 
we have run out of options and reached the empty set. We have chosen to use
$\emptyset$ as the specific set of \states to indicate falsehood in the semantics.
But, as was already pointed out before, any set of inconsistent states would 
have done equally well.

Note that $(i)$ and $(ii)$ have an equivalent pointwise formulation:

\begin{quote}
$(i)$ for each $\tupel{\csp ;\theta}\in S\sem{\phi}$: 
	$\csp\wedge\hat\theta\models_{\cal I}\phi$;

$(ii)$ if $S\sem{\phi}=\emptyset$, then for each 
	$\tupel{\csp ;\theta}\in S$: 
	$\csp\wedge\hat\theta\models_{\cal I}\neg\phi$.
\end{quote}


\begin{lemma}[Preservation]
\mbox{}

\begin{enumerate}
\item  If $\csp \wedge\hat{\theta}\models_{\cal I}\phi_1$ and
$\tupel{\csp ';\theta '}\in \tupel{\csp ;\theta}\sem{\phi_2}$, then
$\csp '\wedge\hat{\theta '}\models_{\cal I}\phi_1$ \hfill{(validity)}
\item If $\csp$, $\hat{\theta}$ and $(\phi_1\wedge\phi_2)$ are consistent
(in ${\cal I}$)
and there is a consistent state $\tupel{\csp ';\theta '}\in 
\tupel{\csp ;\theta}\sem{\phi_2}$, 
then there is a state $\tupel{\csp '';\theta ''}\in 
\tupel{\csp ;\theta}\sem{\phi_2}$, such that
$\csp ''$, $\hat{\theta ''}$ and
$(\phi_1\wedge\phi_2)$ are consistent
(in ${\cal I}$).
\hfill{(consistency)}
\end{enumerate}

\end{lemma}

The first part of the lemma insists that the computation of
$\phi_2$ preserves the validity of $\phi_1$ and that the second part
of the lemma insists that the computation of $\phi_2$ preserves the
consistency of $\phi_1$ (with $\phi_2$ in a suitable state). 
We see that in the second part we are allowed to make a switch from 
$\tupel{\csp' ;\theta '}$ to $\tupel{\csp '';\theta ''}$. (In the proof this
option is only used in the cases for disjunction.)

The proof of the theorem is a simultaneous induction on the construction of 
$\phi$. Simultaneity is required for the negation case.
In the proof we need the preservation lemma crucially in the case for 
conjunction. The proof of the lemma itself is again a simultaneous
induction, this time on the construction of $\phi_2$.

Both proofs follow the corresponding proofs in \cite{lpar}. So, here we 
feel free to restrict attention to the atomic cases---they show how pointwise
equivalence works---and the conjunction cases---they show the crucial
use of the preservation lemma.\footnote{There is minor divergence from
the formulation in \cite{lpar}: we work with $\csp\wedge\hat\theta$ instead
of $\csp\theta$. This facilitates the proofs marginally, as the reader 
familiar with \cite{lpar} can check for himself.}

\begin{proof}[Soundness]

{\bf atoms}\mbox{}

In case $\phi$ is an atomic formula $A$ say:

$S\sem{A}=\bigcup\verz{
\tupel{\csp ;\theta}\sem{A}:\; 
\tupel{\csp;\theta}\in S}
=\bigcup\verz{\infer(\tupel{\csp \cup\verz{A} ;\theta}):\; 
\tupel{\csp;\theta}\in S}$.
\begin{enumerate}
\item Consider $\tupel{\csp ';\theta '}\in 
\infer(\tupel{\csp \cup\verz{A} ;\theta})$. Now: $(\csp\wedge{A})\wedge
\hat{\theta}
\models_{\cal I} A$. So, by pointwise equivalence also: 
$\csp '\wedge\hat{\theta '}\models_{\cal I } A$. So, $\bigvee S\sem{A}
\models_{\cal I} A$, as required.
\item 
%
%
%
Suppose $\infer(\tupel{\csp \cup\verz{A} ;\theta})$ contains only inconsistent
states (for all $\tupel{\csp ;\theta}\in S$). 
By pointwise equivalence we may conclude $\csp\wedge A\wedge\hat\theta
\models_{\cal I} \bot$ for all $\tupel{\csp;\theta}\in S$.
Hence $\csp\wedge \hat\theta\models_{\cal I} \neg A$ 
for all $\tupel{\csp;\theta}\in S$.
From this we conclude: $\bigvee S\models_{\cal I}\neg A$, as required.
\end{enumerate}

Note how pointwise equivalence ensures that $\tupel{\csp ';\theta '}$ has 
an ancestor. Here this can only be one state: $\tupel{\csp\cup \verz{A};\theta}$.\\

{\bf conjunction}\mbox{}

In case $\phi$ is a conjunction, $\phi_1\wedge\phi_2$ say,
$\tupel{\csp ''; \theta ''}\in\tupel{\csp ;\theta}\sem{\phi} $ iff

$\tupel{\csp ''; \theta ''}\in\tupel{\csp ';\theta '}\sem{\phi_2} $  for
some
$\tupel{\csp '; \theta '}\in\tupel{\csp ;\theta }\sem{\phi_1} $.

\begin{enumerate}
\item Let $\tupel{\csp;\theta}\in S$. Then
the induction hypothesis gives:
$\csp ''\wedge\hat{\theta ''}\models_{\cal I}\phi_2$
and
$\csp '\wedge\hat{\theta '}\models_{\cal I}\phi_1$.
By persistence (i) we may conclude:
$\csp ''\wedge\hat{\theta ''}\models_{\cal I}\phi_1$.
So, first order logic now gives:
$\csp ''\wedge\hat{\theta ''}\models_{\cal I}(\phi_1\wedge\phi_2)$
for each $\tupel{\csp '';\theta ''}\in S\sem{\phi}$.
Hence $\bigvee S\sem{\phi}\models_{\cal I}\phi$, as required.
\item Now we know that $S\sem{\phi_1\wedge\phi_2}$ only contains 
inconsistent states.
So, we have: if $\tupel{\csp ';\theta '}\in\sem{\phi_1}$ is consistent,
then $\tupel{\csp ';\theta '}\sem{\phi_2}$ only contains inconsistent states. 
From this we may conclude by induction hypothesis that: for each consistent
$\tupel{\csp ';\theta '}\in S\sem{\phi_1}$,
$\csp '\wedge\hat{\theta '}\models_{\cal I}\neg \phi_2$ ($\otimes$).

Now assume that for some
$[\overline{b}]$:
$\models_{\cal I} (\csp\wedge\hat{\theta}\wedge\phi_1\wedge\phi_2)
[\overline{b}]$
and that we have a consistent
$\tupel{\csp ';\theta '}\in S\sem{\phi_1}$.
Then preservation (ii) tells us that the consistency is preserved, i.e.,
there is a state 
$\tupel{\csp '';\theta ''}\in S\sem{\phi_1}$ and values
$[\overline{b''}]$:
such that:
$\models_{\cal I}
(\csp ''\wedge\hat{\theta ''}\wedge\phi_1\wedge\phi_2)
[\overline{b''}]$.
But this contradicts ($\otimes$). So, for no 
$[\overline{b}]$:
$\models_{\cal I} (\csp\wedge\hat{\theta}\wedge\phi_1\wedge\phi_2)
[\overline{b}]$.
Hence $\csp\wedge\hat{\theta}\models_{\cal I}
\neg (\phi_1\wedge\phi_2)$ (for all $\tupel{\csp ;\theta}\in S$),
which is as required.
\end{enumerate}
\end{proof}

\begin{proof}[Preservation]

{\bf atoms}\mbox{}

In the atomic case $\phi_2 = A$ for some atom $A$ and 
$\tupel{\csp ';\theta '}\in\infer (\tupel{\csp\cup\verz{A};\theta})$.

\begin{enumerate}
\item We know that: $\csp\wedge\hat{\theta}\models_{\cal I}\phi_1$.
So, also: 
$(\csp\wedge{A})\wedge\hat{\theta}\models_{\cal I}\phi_1$.

By pointwise equivalence this gives:
$\csp '\wedge\hat{\theta '}\models_{\cal I}\phi_1$,
as required.
\item
The assumption gives us values 
$[\overline{b}]$:

$\models_{\cal I} (\csp \wedge\hat\theta \wedge (\phi_1\wedge A))
[\overline{b}]$.

It is harmless to add a copy of $A$ to get: 
$\models_{\cal I} (\csp\wedge{A} \wedge\hat{\theta} \wedge (\phi_1\wedge{A}))
[\overline{b}]$.

Now pointwise equivalence ensures that there are 
$[\overline{b'}]$ such that:

$\models_{\cal I} (\csp ' \wedge\hat{\theta '}\wedge (\phi_1\wedge A))
[\overline{b'}]$. 
\end{enumerate}

{\bf conjunction}\mbox{}

In this case $\phi_2=(\psi_1\wedge\psi_2)$ and $\tupel{\csp '';\theta ''}\in
\tupel{\csp ';\theta '}\sem{\psi_2}$ for some 

$\tupel{\csp ';\theta '}\in \tupel{\csp ;\theta}\sem{\psi_1}$. 

\begin{enumerate}
\item 
By induction hypothesis (for $\phi_1$ and $\psi_1$):
$\csp '\wedge\hat{\theta '}\models_{\cal I}\phi_1$.

By a second application of the induction hypothesis
(to $\phi_1$ and $\psi_2$):

$\csp ''\wedge\hat{\theta ''}\models_{\cal I}\phi_1$.
\item 
By assumption:
$\models_{\cal I} (\csp\wedge 
\hat{\theta} \wedge(\phi_1\wedge (\psi_1\wedge\psi_2 )))
[\overline{b}]$.

So:
$\models_{\cal I} (\csp\wedge\hat{\theta} \wedge(\phi_1\wedge\psi_1))
[\overline{b}]$.

By induction hypothesis we get:
$\models_{\cal I} (\csp ''\wedge\hat{\theta ''}\wedge(\phi_1\wedge \psi_1))
[\overline{b''}]$ (for suitable $\tupel{\csp '';\theta ''}$).

Next the induction hypothesis (for $\phi_1\wedge\psi_1$ and $\psi_2$)
provides:

$\models_{\cal I} (\csp '''\wedge\hat{\theta '''}\wedge
((\phi_1\wedge \psi_1)\wedge\psi_2))
[\overline{b'''}]$ (for suitable $\tupel{\csp ''';\theta '''}$,
as required.
\end{enumerate}

\end{proof}

This establishes that all settings of the {\it infer}-parameter that satisfy
the conditions discussed in section \ref{cond}, result in sound semantics for 
first order logic: all the instantiations of {\it infer} only produce
outcomes of $S\sem{\phi}$ that satisfy $\phi$ and only report {\it false}
if $\phi$ is false in $S$. In \cite{lpar} we show how a large number
of forms of constraint propagation are sound instances of the 
{\it infer}-parameter. It is clear that also lots of search tricks can
be modeled as settings of the {\it infer}-parameters. (See section \ref{conc}
for the discussion of our motivating examples.) If these settings of {\it infer}
obey the pointwise equivalence condition, they will lead to a sound
instantiation of the semantics. So, we can now also read the conditions in section \ref{cond} as {\it conditions for sound search}.

\section{Looking back and ahead}
\label{conc}

Let's go back to the motivating examples from section \ref{intro}.
There we used the examples to illustrate the use of all kinds
of `disjunctive splits' of states into substates to model search 
and subproblem selection. Now it is clear that such disjunctive splits
leads to definitions of the {\it infer} parameter that
obey the new, {\it pointwise} equivalence condition. For example, 
if $\verz{x=1\vee x=-1;\; \epsilon}$, then $\verz{\tupel{x=1;\; \epsilon},
\tupel{x=-1;\; \epsilon}}\subseteq\infer S$
is consistent with the conditions on {\it infer} that we propose.  
Similarly, $S=\verz{x\in D_x;\; \epsilon}$ and $D_x=D_{good}\cup D_{bad}$, 
then $=\verz{\tupel{x\in D_{good};\; \epsilon},
\tupel{x\in D_{bad}\; \epsilon}}\subseteq \infer S$ satisfies pointwise equivalence.
So, the adaptation of the definitions pays off: the system proposed in 
\cite{lpar} has now been extended to include the investigation 
of such search strategies in a sound way.

This means that we now have an extremely rich system:
\begin{itemize}
\item the denotational 
semantics for first order logic that we present gives natural computational 
readings for the logic connectives (following \cite{Apt00}); 
\item it allows for
the investigation of a wide variety of forms of constraint propagation
(following \cite{lpar}); 
\item and now it also includes the option of analyzing
search routines (as suggested in \cite{WJ02}). 
\end{itemize}
All these ingredients 
are combined in one system in such a way that {\it soundness} with respect
to the standard interpretation of first order logic is preserved. Hence we 
can regard the conditions on {\it infer} that we have presented as conditions 
for {\it sound search} in constraint programming.

The soundness theorem shows how attractive the combination of ingredients
proposed is for establishing general results. In \cite{lpar} (and here in
section \ref{recol}) it was shown how different forms of constraint 
propagation can be seen as instantiations of the {\it infer} parameter.
Hence these forms of constraint programming can be covered all at once, 
by proving one theorem only.
Here we extend the level of generality to several ideas about search.
Of the two examples of search in this paper,
the first example has didactic 
merits only: it certainly is not a `hot issue' in the current literature.
But this first example
indicates in a convincing way how other sorts of search tricks also fall
within the scope of our proposal. In particular the search tricks based
on ranking of values of variables from \cite{WJ02}. Such search tricks 
certainly are 
a real issue in the current literature on constraint programming. And we can
cover them in the proposed analysis.

We have given a soundness theorem as an example of general results.
Soundness is a natural requirement on search techniques: we 
do not want to lose anything as we are searching. And the way in which we 
translate this soundness claim into the format of \cite{Apt00} is extremely
natural. But we do not yet give an
equally natural way of translating other hot issues concerning search, such as
efficiency claims about search tricks,
into the format. This a clearly an interesting task for future research.
As a starting point for such investigations we see \cite{deci}.
There the axiomatization and decidability of the denotational semantics in 
Apt's \cite{Apt00} is discussed in 
a way that allows us to estimate upperbounds for the
complexity of the semantics. If these results are combined with conditions
on the complexity of the {\it infer} parameter, this should allow the 
analysis of efficient combinations of computation, constraint
propagation and search.


\end{document}